# A Collaboration Network Model Of Cytokine-Protein Network

Sheng-Rong Zou, Ta Zhou, Yu-Jing Peng,

Zhong-Wei Guo，Chang-gui Gu, Da-Ren He*

Yangzhou University, Yangzhou 225002, P. R. China

Abstract：Complex networks provide us a new view for investigation of complex systems. In this paper we collect data through STRING database and present an cytokine-protein network model with cooperation network theory. The results show that this model could describe the cytokine-protein collaboration preferably．The basic attributes of this network is well consistent to our preknowledge of cytokine-protein system．

Keywords：complex network, cytokine-protein interaction，immune system，collaboration network, bioinformatics

## 1. Introduction

The study of networks has exploded across the academic spectrum in the past some years. In particular, in biology network studies, it is increasingly recognized the role played by the cellular networks, the intricate web of interaction among genes, proteins and other molecules regulating cell activity, in unveiling the function and the evolution of living organisms. It is interesting to investigate the network descriptions on cytokine-protein system .

Although many details of particular cytokine interactions have been elucidated and the effects of cytokines on a myriad of cellular functions have been described, practically nothing is known about the behavior of the system as a whole. All cytokine collaboration exhibit nonlinear behaviors. In fact, they act in a complex, intermingled network where one cytokine can influence the production of, and response to, many other cytokines. So we believe that cytokine network should be more effectively described by a complex network.

Because of the complexity of cytokine network, it is necessary to study ulteriorly the characteristic of single cytokine. We can comprehend the characteristic of cytokine through the cytokine-protein interaction.

The paper is organized as follows. In section two we shall present our method and model. In section three we shall present some results. In the last section, we make a conclusion.

## 2. our method and model

Other authors have modeled the cytokine-protein model, with a variety of approaches and areas of emphasis. But many essential features of this complex system are still not understood. We collect data through string database and construct a cytokine-protein network with String database and the bipartite graph theory.

### 2.1 generalized collaboration network

Here we use the method of collaboration network. We first introduce the knowledge of collaboration network. There have been considerable of interest in the study of a special class of social networks, called social collaboration networks. These include movie actor collaboration networks and scientist collaboration networks. This kind of networks can be described using bigraph(bipartite graphs) as shown in Fig. 1. One type of nodes can be called "actor" such as movie actors or scientists, which are shown in the bottom row, indexed with $L_i$; $i = 1, 2, \ldots$. Another type of nodes can be called "act" such as movies or scientific papers, which are shown in the top row ($U_i$; $i = 1, 2, \ldots$). In these graphs, only undirected edges between different types of nodes are considered.

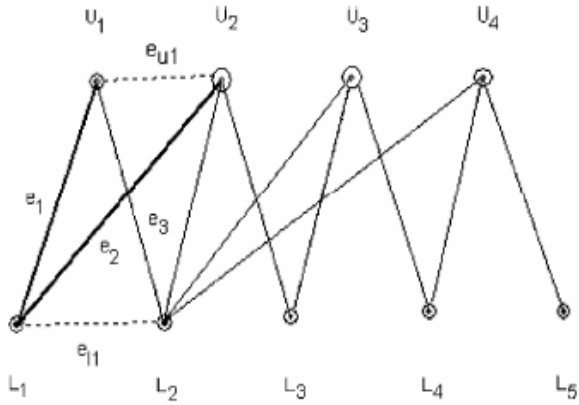

Fig. 1. A bi-partite graph for describing collaboration networks.

We draw them in Fig. 1 with a solid line $e_i$; $i = 1, 2, \ldots$. edge represents an actor taking part in an act. If we consider one type of nodes only, two edges sharing a common vertex in the bigraph are projected onto an edge between the two nodes of the same type. Take, for example, the movie actor collaboration network. Sometimes, we need to consider only the collaboration between actors. In this situation, edge $e_{l1}$ between $L_1$ and $L_2$ in Fig. 1, which is obtained by projecting $e_1$ and $e_3$ to the bottom row, shows their collaboration in the same film $U_1$. If two actors cooperate in more than one film, the relation can be expressed by multiple edges between them. On the other hand, we can define an edge between two films (two $U_i$ vertices), which indicates that the same common actor takes part in both films. The edge $e_{u1}$ between $U_1$ and $U_2$ in Fig. 1, which is a projection of $e_1$ and $e_2$ to the top row, indicates that $L_1$ takes part in these two films. If two films involve more than one common actor, the relation can also be expressed by multiple edges. The larger the number of edges between two films is, the more similar characteristics these films share. Newman and Li used connection weights to denote multiple edges, and studied the resulting weighted networks in Refs.. But we prefer to retain multiple edges to make our model clearer. To consider how many films actor $i$ acted in all, we define a quantity $h_i$, ''act-degree of actor' I', which is equal to the number of U nodes linked to $L_i$ in the bigraph, such as the four thin lines emitted from $L_2$ in Fig. 1. Obviously, the four U nodes $U_1$; $U_2$; $U_3$ and $U_4$ form a complete graph in the up-projected graph consisting of only U nodes. Similarly, if we have to consider how many actors are taking part in film j, we can define a quantity Tj, "act-size", which stands for the number of actors in act j, and it is equal to the number of L nodes linked by the node Uj in the bigraph. Again, these L nodes form a complete graph in the down-projected graph consisting of only L nodes. Each node has a degree value Tj _ 1. Of course, two complete graphs may share one or more edges in the down-projected graph, however, the conclusion that the degree of each node equals Tj _ 1 still holds when multiple edges are present. If we extend the concept of a "complete graph" to the situation where multiple edges are considered, i.e., define a graph where each pair of nodes are connected by edges (including multiple edges) as a complete graph, it is easy to verify that such a down-projected network is still a set of complete graphs.

**2.2 our model**

The cytokine-protein network we consider is constituted by two kinds of nodes, one is immune protein types, which can act as actors; another is cytokine, which can act as act. The cytokine-protein system model we construct with 216 act, 62994 actor, 5391 links. These data are small compared to classic social collaboration networks such as the Hollywood actors collaboration network and scientists collaboration network, but it represents a network of very different origin.

**3. Results**

**3.1 Accumulative act degree distribution and act degree**

The accumulative act degree distribution and act degree is an important geometric properties of the complex network. Accumulative act degree (multiple edges are counted) is that an actor take part in how many act. Fig 2 shows the accumulative act degree distribution (with multiple edges counted) of the cytokine-protein system. The distribution can be well described using SPL[3].

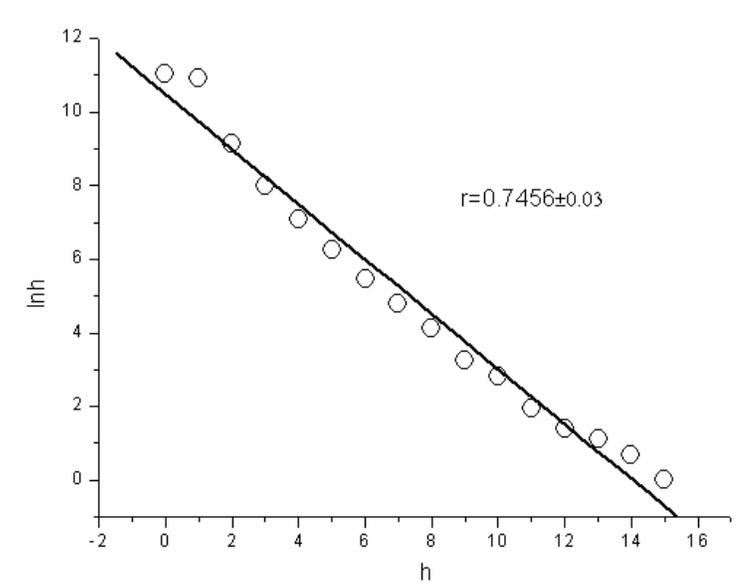

Fig 2  The accumulative act degree distribution

Fig 2 The accumulative act degree distribution (multiple edges are counted) of cytokine-protein system. The inset shows the results of an accumulative act-degree distribution.

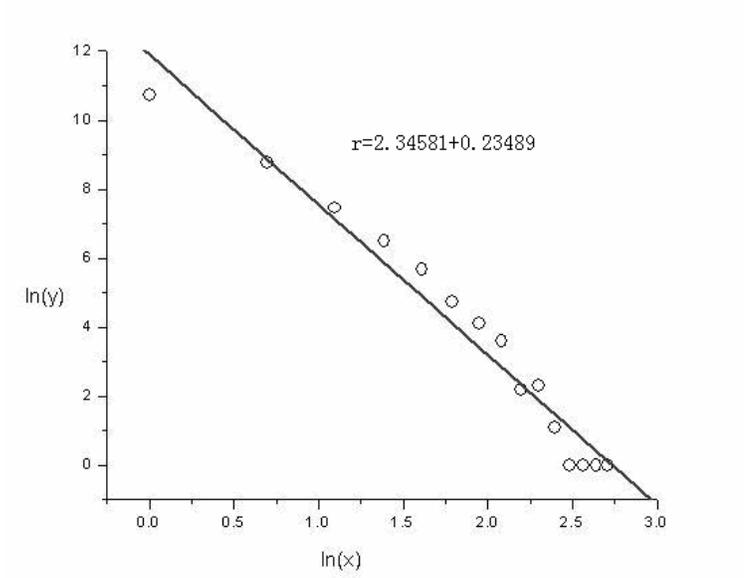

Fig 3  Act degree distribution

Fig 3  Act degree is that an actor take part in how many act.  Fig 3 shows the act degree of the cytokine-protein system, the distribution can be well described using power –law with index of 1.8. It reveals that act degree linear preference rule is used in cytokine collaboration of cytokine-protein system.

At the same time, we calculate the average shortest path of cytokine-protein system. The formula shows the results of path and path connect. The length of the average shortest path is 1.29. The formula as follows:

$$L = \frac{1}{\frac{1}{2}N(N+1)} \sum_{j \neq i} d_{ij}$$

N is the node of the system; $d_{ij}$ is the short path length between one node and another node.

### 3.2 Accumulative dyad act degree and dyad act degree

Accumulative dyad act degree and dyad act degree are another important properties to generalized collaboration network.. Dyad is that two actors and the relationship of them. Accumulative dyad act degree (multiple edges are counted) is that a dyad take part in how many act. Fig 2 and Fig 3 respectively shows the accumulative dyad act degree distribution and dyad act degree distribution of the cytokine-protein system, the distribution can be well described by typical SPL[3] functions. It reveals that dyad linear preference rule is used in cytokine-protein system. Construction act that is not only the most common use of dyad, also considered a variety of specific factors to each dyad,

Often because there are too many specific factors and the lack of interconnected. It is equal to random selection. So SPL dyad act degree distribution rules should be the common features of the most generalized network of cooperation. In our cytokine-protein system, dyad act degree can be said for any two mediators in a cell for the probability of successful cooperation.

### 4. Conclusion

From our act degree distribution, we find that HRAS and TNFRSF13C are highly collaborated with other proteins. S100A8、S100A1、MAPK8、S100A7、LIF、CCL4、 a CXCL13 are relatively highly collaborated with other proteins. It reveals these proteins. are important in cytokine-protein system to regulate their cytokine activity. We also find that LIF and HRAS are important protein in cytokine-protein system. The result is well consistent to our preknowledge of cytokine-protein system.

## Acknowledgements

The research is supported by National Natural Science Foundation of China under the grant No. 70671089 and 10635040.